\documentclass[9pt]{sigplanconf}

\usepackage{amsmath}
\usepackage{epsfig}
\usepackage{algorithmic}
\usepackage{algorithm}

\begin{document}

\conferenceinfo{CCEM '12}{Oct 11-12, Bangalore, India.} 
\copyrightyear{2012} 
\copyrightdata{[to be supplied]} 


\title{Building Resilient Cloud Over Unreliable Commodity Infrastructure}

\authorinfo{Piyus Kedia, Sorav Bansal}{IIT Delhi}{}
\authorinfo{Deepak Deshpande, Sreekanth Iyer}{IBM India Software Lab}{}
\maketitle

\begin{abstract}
Cloud Computing has emerged as a successful computing paradigm for efficiently utilizing managed compute infrastructure such as high speed rack-mounted servers, connected with high speed networking, and reliable storage. Usually such infrastructure is dedicated, physically secured and has reliable power and networking infrastructure. However, much of our idle compute capacity is present in unmanaged infrastructure like idle desktops, lab machines, physically distant server machines, and laptops. We present a scheme to utilize this idle compute capacity on a best-effort basis and provide high availability even in face of failure of individual components or facilities.

We run virtual machines on the commodity infrastructure and present a cloud interface to our end users. The primary challenge is to maintain availability in the presence of node failures, network failures, and power failures. We run multiple copies of a Virtual Machine (VM) redundantly on geographically dispersed physical machines to achieve availability. If one of the running copies of a VM fails, we seamlessly switchover to another running copy. We use Virtual Machine Record/Replay capability to implement this redundancy and switchover. In current progress, we have implemented VM Record/Replay for uniprocessor machines over Linux/KVM and are currently working on VM Record/Replay on shared-memory multiprocessor machines. We report initial experimental results based on our implementation.
\end{abstract}


\section{Introduction}
\label{sec:intro}
Virtualization of server resources is the building block of cloud computing.
Almost all cloud installations use dedicated high-performance server hardware
resources with reliable and secure power, cooling and network infrastructure.
Such installations are expensive, both at setup and operationally.
On the other hand, much compute capacity in the form of idle desktops, lab
machines, small server clusters, and other office computing equipment, remains
underutilized. Under-utilization could be perennial (e.g., a machine that
gets used
mostly for lightweight browsing and/or work processing) or
time-dependent (e.g., idle at night time or during summer months). Such
scenarios are especially common in academic institutions (computing
lab infrastructure, small high-performance compute clusters for individual
research groups), software dev/test environments (multiple desktops
per user, or servers used to run builds and automated tests), and organizations
with infrastructure across the world with different time zones.
We attempt
to utilize this unused capacity to support a cloud
service, thus reducing wastage and lowering costs.

We run IaaS (infrastructure-as-a-service) cloud on these underutilized
computers. The cloud virtual machines (VMs) are run on the
underutilized computers in
the background. We use hypervisor support on the host operating system to
run these VMs without disrupting other ongoing activity. The cloud VM
resource usage is maintained low enough to avoid any visible performance
effects to the end-user of that computer. VMs are dynamically scheduled
and migrated among the available computers appropriately, taking the
underlying physical network topology into consideration. We call this
model, {\em community cloud computing}, as it requires a community
of users to allow usage of their underutilized resources.
Because we use shared and relatively lower performance hardware to run
the VMs, we do not expect to run performance-critical
workloads on the community cloud. We expect the model to
best suit long-running
workloads which are less performance critical. e.g., compute-intensive
scientific workloads or data-intensive analytic workloads. 
Such non-performance-critical workloads are often run on
time-shared hardware in managed cloud environments to lower
end-user costs; running
them on
community hardware (with less sharing) could provide similar or better
performance at the same cost.
Rough estimates
of the underutilized compute capacity available in such unmanaged
infrastructures indicate that this model, if successful, could
result in significant cost savings.

As far as we can see, the primary challenge in achieving a successful
implementation of a community cloud is maintaining reliability
and availability, given
that most of our physical infrastructure is unreliable. The user of
a community cloud should expect availability
similar to that on a regular cloud. In fact, migration of VMs between
managed and unmanaged infrastructure should be opaque to the
user --- the user should simply expect the same level of performance and
reliability irrespective of where the VM is running. Community-managed
computers have unreliable power (no redundancy in power supplies, poor
UPS support, users unplugging power cables in error, etc.), unreliable
network connectivity (faulty network wires/NICs/hubs, users unplugging
network cables, routers losing power supply, etc.), and unreliable
storage (failing local disks, unreliable connections to NFS storage, etc.),
among other things.

We propose the use of Virtual Machine Record/Replay
to implement redundancy of live VM
images (a similar approach is also used to implement high-availability
services in commercial settings\cite{vmware_ha}). Each VM is
simultaneously
executed on $n$ different physical machines, where $n$ is the degree of
redundancy. One of the VM replicas is designated the primary and executes
on live network state, while the other $n-1$ (secondary) replicas simply
replay the primary. On failure of the primary, one of the secondary replicas
takes over. Because, the secondary replica was replaying the primary, its
execution state is expected to be identical (or very close) to that of the
primary at failure
time. The number of secondary replicas can be always maintained at
$n-1$ (barring the short intervals of time after failure of either primary
or secondary replicas, when a new secondary needs to be spawned). The choice
of the different physical machines which execute
the replicas of one VM can be made taking the network and power supply topology
into consideration (e.g., failure of one network router should not cause
all replicas to become disconnected). We expect a redundancy factor
of 3-5 ($n$=3-5) to provide
acceptable reliability for our cloud service.

There are two primary challenges in realizing a practical implementation
of our model:
\begin{enumerate}
\item {\em Failover based on record/replay}: We need an efficient
record/replay and failover mechanism. We implement uniprocessor record/replay
on Linux/KVM and discuss its performance in detail. The performance of our
implementation is comparable to that of VMware's commercial closed-source
workstation product\cite{vmware:workstation}. For seamless network
operations on failover, we require that all VM replicas belong to the same
VLAN. On primary failure, a secondary replica becomes the primary and
the network routing is automatically adjusted using ARP network
protocol. At failover time, the secondary could be slightly behind the
primary (e.g., due to network latency of streaming the record log), and this
could result in application-level inconsistencies over the network. Fortunately,
we find that this happens rarely in practice. We discuss our
failover experiments and findings in detail.
\item {\em Scheduling}: The replicas of multiple VMs need to be 
scheduled (placed) across multiple physical machines with the following
constraints:
\begin{itemize}
\item VMs should only be run on under-utilized hosts. If a host shows high
utilization, the VM should be migrated to another host.
\item Different replicas should be placed across different physical machines
and preferably on different physical network branches (to guard against
router failures)
\item The VM image store should always be available to all VM replicas. This
could be either done using a reliable NAS server or using a distributed
(and replicated) filesystem. With a NAS server, VM placement is further
constrained to be on
hosts close to the storage, to minimize network traffic. For a distributed
filesystem using commodity disks on the community computers, efficient
algorithms need to be devised to place the storage redundantly such that
it guards against failure and still remains accessible to the computing hosts
without too much network traffic.
\item Finally, the replica placement must remain sensitive to the network
traffic caused due to the streaming of record logs between primary and
secondaries.
\end{itemize}
These scheduling constraints are difficult, and require significant
experimentation. We have not yet fully explored the solution space to this
problem in this paper. We provide an initial assessment of the tradeoffs
involved, and intend to study them experimentally in detail in future.
\end{enumerate}

There are other alternatives to VM record/replay for implementing redundancy,
such as periodic snapshotting of VM state or file system record/replay.
The choice of approach depends on the tradeoff between the performance
overhead during normal operation, and the nature of state loss on failure.
VM Record/Replay operates on 5-10\% runtime overhead while maintaining a
loss of typically a few 100 milliseconds of execution state. On the other
hand, file system record/replay techniques have lower runtime overhead but
require a full VM restart/reboot on failure (resulting in an execution state
loss of 10s of seconds). Periodic snapshotting of VM state also suffers from
the drawback of loss of large execution state on failure. We believe that
record/replay has the best performance and reliability tradeoffs among the
available alternatives.

The paper is organized as follows. Section~\ref{sec:record_replay} discusses
our implementation of VM record/replay and failover on Linux/KVM.
Section~\ref{sec:results} discusses
our experiments and results, Section~\ref{sec:relwork} discusses related
work, and Section~\ref{sec:conclusion} concludes.
\section{Virtual Machine Record/Replay}
\label{sec:record_replay}
ReTrace\cite{retrace} demonstrated the capability to record/replay (R/R)
an execution
in VMware Workstation
and reported as low as 5\% runtime overhead, and 0.5 byte per thousand
instructions log growth rate. VM R/R works by recording all
external input to virtualized devices and the timing of interrupts
as they are delivered to the guest. Time is counted by counting
the number of instructions (or branches) executed by the
guest. On x86 platforms, the
three tuple {\tt (nbranches, rip, rcx)} uniquely identifies a
logical guest execution epoch, where {\tt nbranches} is the number
of branches executed by the guest, {\tt rip} is the guest's
current instruction
pointer,
and {\tt rcx} is the current value of count register (needed for
instructions with {\tt rep} prefix). The {\tt nbranches} counter
is maintained using hardware performance counters.
All deterministic
instructions (i.e., instructions producing identical output
on same input irrespective of time of execution) can be executed unmodified
directly on hardware. All non-deterministic instructions must be made
to trap to the Virtual Machine Monitor (VMM) and their non-deterministic result
recorded (e.g., {\tt rdtsc} on x86).
Because a huge fraction
of executed instructions in common workloads are deterministic
(e.g., $>$99\% for most compute-intensive workloads), the overhead
of VM R/R is minimal. All interrupts delivered to the
guest are recorded alongwith their epoch {\tt (nbranches, rip, rcx)}.
Emulated devices are recorded by recording all the non-determinism
in the device emulation code. For example, if the device uses an
external input (e.g., network packets), that input is logged.

During replay, the results of non-deterministic instructions and
non-deterministic device inputs are obtained from the log.
Replaying interrupts requires special care. We require the guest
to yield control to the VMM at the interrupt epoch
{\tt (nbranches, rip, rcx)} for accurate delivery of the replayed
interrupt. On x86 architectures, this can be achieved
by configuring the performance counters to overflow at the
desired branch count {\tt nbranches}, which generates an interrupt
causing the guest to yield control to the VMM. We then single-step
the guest till we reach the desired {\tt (rip, rcx)} before injecting
the replayed interrupt to the guest.
On current x86 architectures, this interrupt-on-overflow mechanism
for performance counters is imprecise.
It is possible for the branch count to overshoot
the desired value by up to 128 before an interrupt is generated. The solution
to this problem (as also noted in \cite{smprevirt}) is to configure the
performance counters to generate an
interrupt at {\tt (nbranches - 128)} and then single-step the guest
till the interrupt injection epoch {\tt (nbranches, rip, rcx)}. This
careful singlestepping near interrupt injections causes extra runtime
overhead during replay (compared to record).

There are two alternatives while replaying the virtual disk device, namely
{\em full-replayed disk} or {\em output-replayed disk}. A full-replayed disk
is snapshotted at the start of the recording session so that
its state remains reproducible at any execution epoch. The reads/writes
to the disk do not need to be recorded, as they will be identical during
record and replay, provided all other external and timing
related non-deterministic
inputs remain identical. For an output-replayed disk, the disk device
is not snapshotted and only the values returned by the disk on
every disk read are recorded. Here, the disk state cannot be reconstructed
but the CPU/memory state can still be reconstructed at any execution
epoch. We use full-replayed disks for our experiments.

We implement redundancy using one primary and multiple secondary VM replicas,
all running
on different physical hosts simultaneously. The primary records and
all secondaries replay. The record log (also called execution trace)
is
streamed from the primary
to the secondary using TCP/IP.
On failure, we switch primary to one of the secondaries.

We discuss the performance
of our VM R/R implementation and the
failover mechanism in the next section.

\section{Experimental Results}
\label{sec:results}
We implemented KVM R/R inside Linux
kernel 2.6.36.4.
Our implementation uses x86 hardware performance
counters to count branches, and the monitor trap flag (MTF)
for single-stepping. We implemented record/replay functionality
for emulated devices inside Qemu. Our implementation is complete enough
to run a full Linux guest.
We tested our
implementation with all types of applications including graphical and
networked applications. For example, we could successfully view a
Youtube video while running record/replay underneath. We tested
the stability of our implementation by running it continuously for
over 24 hours with an active guest.

We used
Qemu 0.13.0 with default configuration for device emulation record/replay.
Our 32-bit Gentoo Linux guest was configured with fully-emulated e1000
network and IDE disk devices.
The guest ran with 128MB physical memory, 512MB swap space, 8GB disk
space and other devices emulated by Qemu by default.
We ran our experiments on a machine with 12
Intel Xeon X5650 2.67 GHz processor cores, 24GB memory, and 300GB disk.

\begin{figure*}[tb]
\centerline{\epsfig{figure=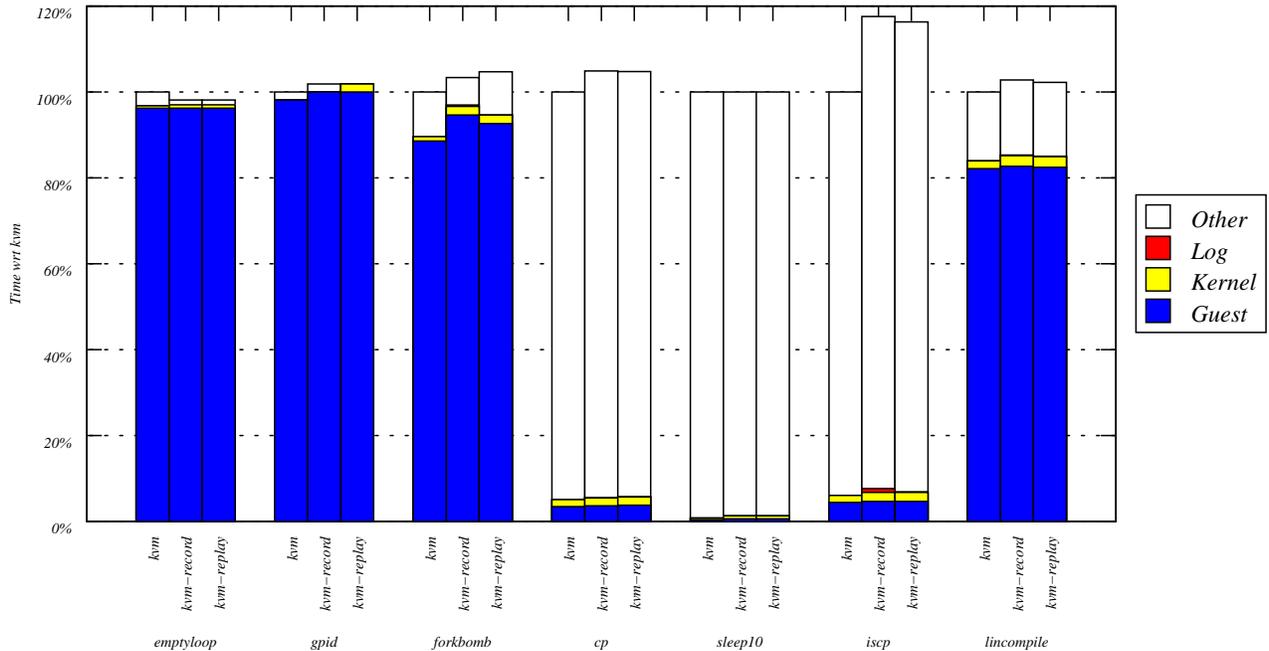, width=2\columnwidth}}
\caption{\label{fig:benchmarks}Performance of KVM Record/Replay.}
\end{figure*}

We evaluate our
implementation using the benchmarks listed in Table~\ref{tab:benchmarks}.
The benchmarks have been chosen to stress different
components of the virtual machine monitor. Some of the benchmarks
have been inspired by a previous VMM
performance study\cite{agesen_vmm_benchmarking}.
The table also presents the typical logsize growth rate of each
benchmark.
The two logsize growth columns are for different
ways of recording the disk device, namely output replay (records and
replays the output of the disk device) and full replay (emulates the
disk device fully). Figure~\ref{fig:benchmarks} shows
the performance characteristics of our VM R/R implementation on KVM.
We show results compared to KVM as baseline. We deliberately don't show
results compared to native execution, because performance difference between
KVM and native execution is either negligible (for compute-intensive
workloads) or is heavily
dependent on the chosen virtual devices (for I/O-intensive workloads). We
simply use the default KVM/Qemu virtual devices.
\begin{table*}
  \begin{center}
  \begin{tabular}{l|l|c|c}
    Benchmark & \multicolumn{1}{|c|}{Description} & Log Growth (full) & Log Growth (output)\\
              &             &     KB/sec     &     KB/sec  \\
    \hline
    {\tt emptyloop} & A process running a compute-intensive loop & 32.3 & 32.5 \\
    {\tt gpid} & A process repeatedly calling the {\tt getpid} system call on Linux & 33.0 & 33.1 \\
    {\tt forkbomb} & Forks 1 million processes, each of which exits gracefully & 65.8 & 89.2\\
    {\tt cp} & Copies a 100MB file within the same directory on Linux ext3 & 10.6 & 5080\\
    {\tt inet} & Receives data over TCP socket in 4-byte chunks& 793 & 791\\
    {\tt onet} & Sends data over TCP socket in 4-byte chunks & 714 & 732\\
    {\tt sleep} & Calls {\tt sleep(10)} (idles for 10 seconds) & 30.2 & 32.1\\
    {\tt iscp} & Copies a 100MB file from the host machine to the guest VM & 755 & 732 \\
               & using {\tt scp} over the emulated network card at maximum rate & & \\
    {\tt lincompile} & Builds the Linux kernel from source& 42.7 & 195 \\
  \end{tabular}
  \caption{\label{tab:benchmarks}Benchmark description and logsize growth characteristics for full-replayed and output-replayed disks.}
\end{center}
\end{table*}

The {\tt emptyloop} benchmark is compute-intensive and represents all
compute-intensive workloads executing at user privilege level;
{\tt gpid} exercises the system call handling
mechanism in Linux; {\tt forkbomb} exercises the
process creation and destruction methods (including page table manipulations);
{\tt cp} exercises the disk; {\tt inet} and {\tt onet} stress the network;
and {\tt sleep} emulates an idle system. {\tt iscp} exercises both network
and disk, while {\tt lincompile} is a macrobenchmark that combines CPU, memory,
and disk usage. We measure performance
using the host's wall clock time.
Figure~\ref{fig:benchmarks} shows the performance of VM
Record ({\tt kvm-record}) and VM Replay ({\tt kvm-replay}).

We use full-replayed disks for
our R/R experiments.
Each runtime
is divided into the percentage of time spent in the guest, in the
host kernel, and other activities including I/O and idle time.
The overhead of recording is within 20\% for all our benchmarks.
The performance
of VM Replay can be worse due to single-stepping. The overhead
of VM R/R on compute-intensive applications ({\tt emptyloop} and
{\tt gpid}) is almost zero. The overhead is more for I/O intensive
applications such as {\tt cp} and {\tt iscp}. {\tt forkbomb}
has higher R/R overhead than {\tt emptyloop} due to paging activity and swap
disk usage. The
performance of {\tt sleep} on {\tt kvm-replay} is faster than on
{\tt kvm} because executions of the x86 {\tt halt} instruction by the
guest's idle thread finish instantaneously on {\tt kvm-replay}, while
they wait for an external interrupt on {\tt kvm} and other variants.
Compute-intensive applications spend
almost all their time in the guest, while I/O intensive applications
spend a significant fraction of time in the host user-level code
for device emulation.

\begin{figure*}[tb]
\centerline{
\epsfig{figure=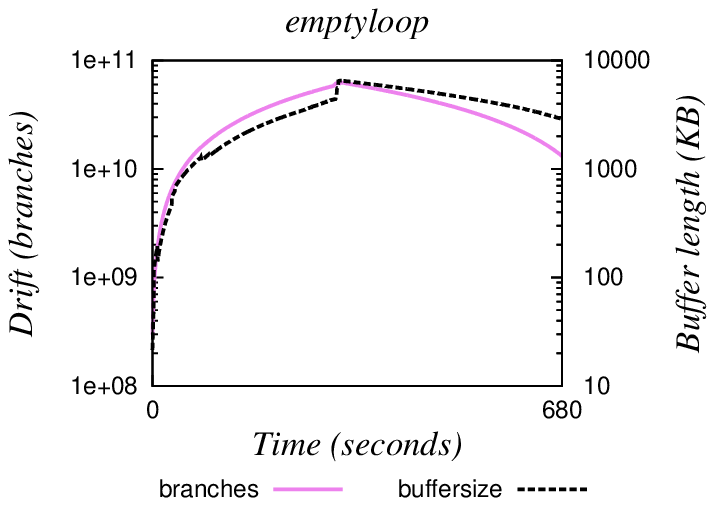, width=0.55\columnwidth}
\epsfig{figure=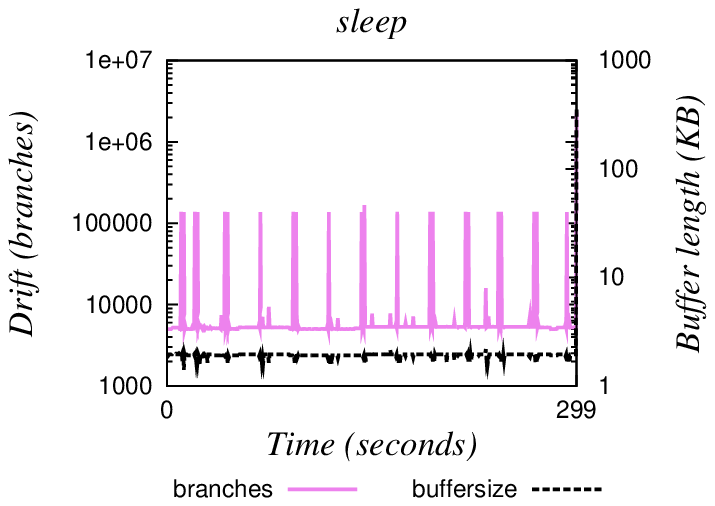, width=0.55\columnwidth}
\epsfig{figure=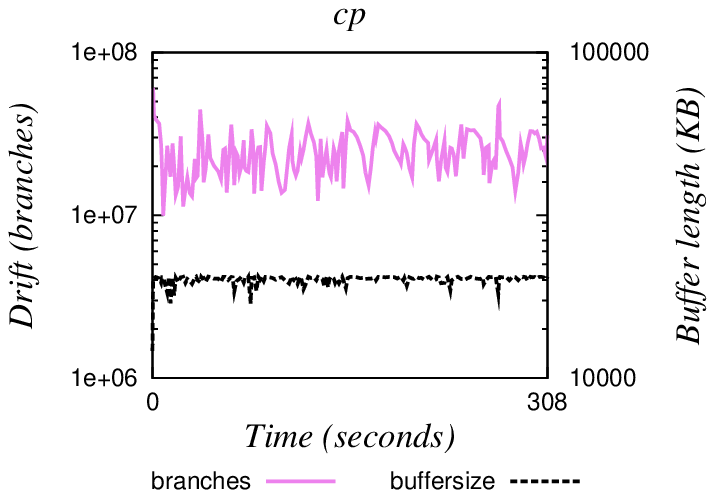, width=0.55\columnwidth}
\epsfig{figure=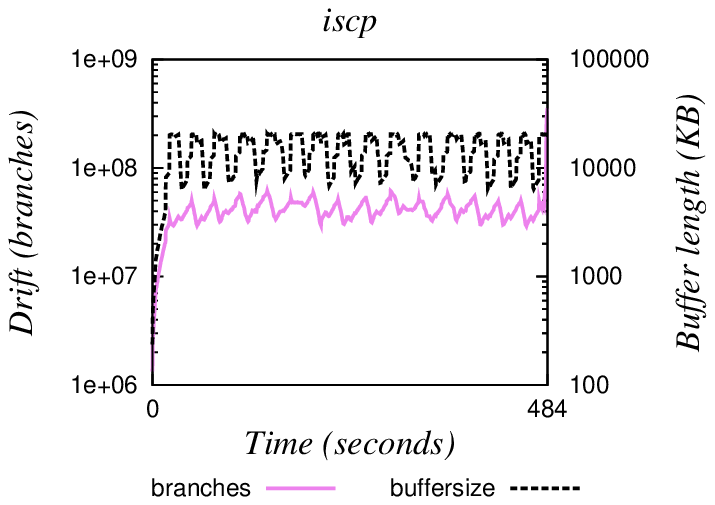, width=0.55\columnwidth}
}
\caption{\label{fig:drift}Distribution of drift across time for {\tt emptyloop}, {\tt sleep}, {\tt cp}, and {\tt iscp}.}
\end{figure*}

Our performance results indicate that recording and replaying incurs only
small overheads
for uniprocessor VMs and it is thus feasible to enable recording on user-facing
cloud VMs. Our results also indicate that replaying could sometimes be
slower than recording. The difference in performance
is due to the hardware support for branch counters and has implications while
running VMs redundantly in primary/secondary mode. If replay is slower,
the replaying process could slowly
{\em drift} from the recording process causing the replaying process to be
much behind after a long time. This drift can be upper-bounded by forcing
the record process to wait for the replay process after the drift exceeds
a limit. This can result in further undesirable loss of performance (in our
experiments, we observe 10-20\% performance loss on compute-intensive
workloads). The performance loss is only seen on compute-intensive workloads.
The drift value also indicates the magnitude of the
``loss of execution state'' in the event of a primary failure, because the
logical (user-visible) state will appear to have rolled back by the drift
value.
Figure~\ref{fig:drift} plots the drift over time for different workloads.
Only {\tt emptyloop} results in high drift values over time.

Failure of the primary replica could potentially result in network-level
inconsistencies due to the rollback in the logical state of the VM.
Because almost all network protocols are resistant to host failures, such
inconsistencies are usually recoverable.
We performed experiments to measure this effect. We ran connection-less
{\tt ping} and {\tt http} servers on our VM and
accessed them through a remote client. We then repeatedly failed the primary
replica and observed that both {\tt ping} and {\tt http} requests continued
after small intervals of unresponsiveness. We next ran a long
connection-oriented
{\tt scp} session ({\tt iscp}) on our VM and caused the primary replica to
fail while the session was active. On failure, the {\tt scp} session stalled.
Analysis of network traffic trace revealed that a rollback caused the TCP
sequence numbers to be rolled back which confused the remote client which
had previously observed packets with higher sequence numbers.

We propose the use of {\em delayed network sends} to deal with this problem.
Here, network packets originating from the primary replica are buffered
(and not {\em released} to the external world) till a secondary replica has not
replayed the send of the network packet. After the secondary replica has
reached the execution epoch of the packet send, the packet is actually
released into the external network. If a failure occurs, the network buffer
is discarded. This ensures that no external inconsistencies occur on
primary failure. We ran {\tt iscp} under this delayed network send configuration
and confirmed that the session continued uninterrupted (although with temporary
performance effects) even on failure.

\begin{figure}[htb]
\centerline{\epsfig{figure=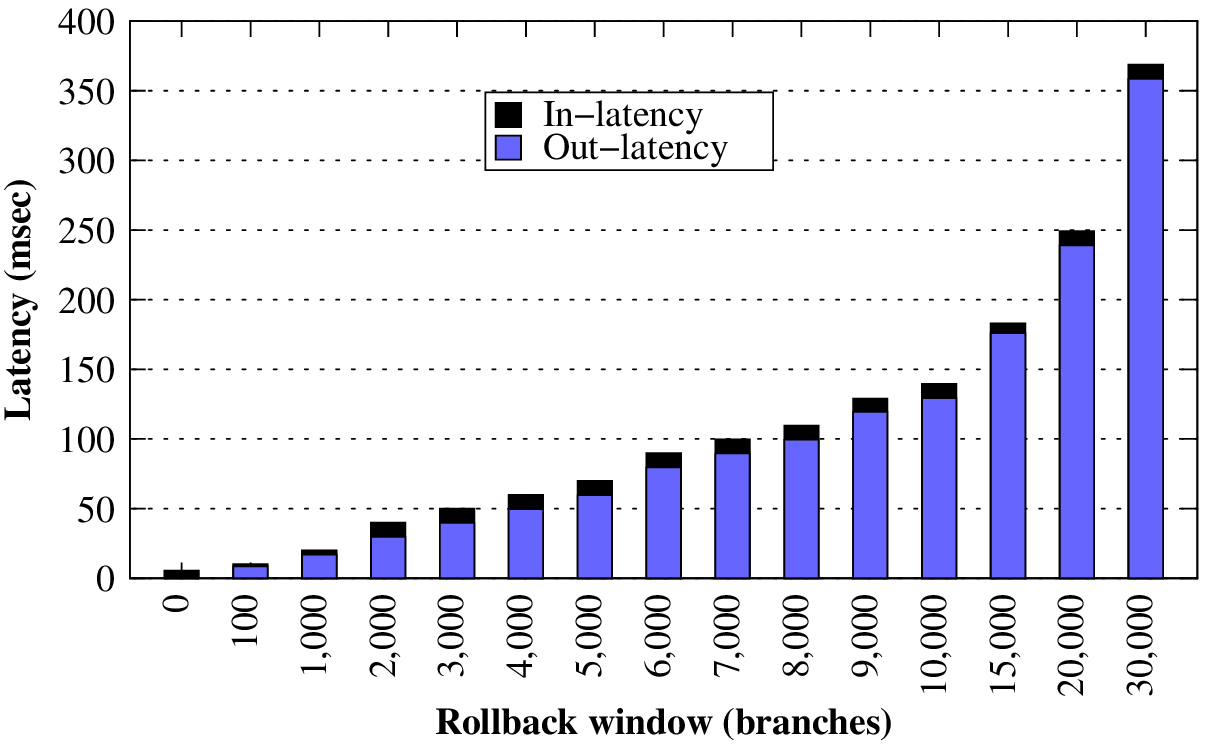, width=\columnwidth}}
\caption{\label{fig:mrnd_latency}Ping latency with increasing rollback window size on {\tt kvm-mrnd}}
\end{figure}

\begin{figure}[htb]
\centerline{\epsfig{figure=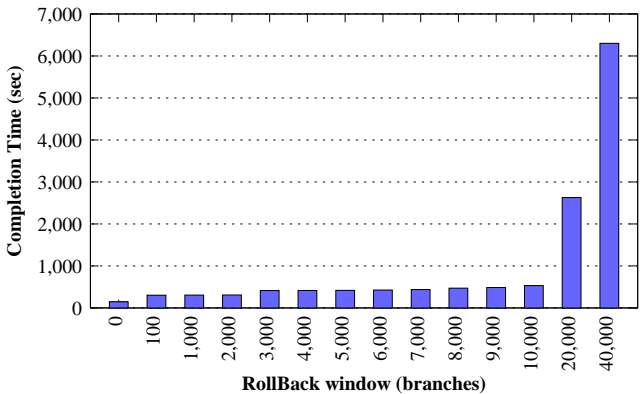, width=\columnwidth}}
\caption{\label{fig:mrnd_iscp}Completion time for {\tt iscp} with increasing rollback window size on {\tt kvm-mrnd}}
\end{figure}

Delayed network sends have effects on the network latency and throughput,
however. To measure this, we configured a primary and a secondary with
delayed network sends and measured latency and bandwidth with increasing
drift between the primary and secondary. We measure drift using the number
of branches executed by the guest, which is an indirect but more deterministic
measure of time.
Figures~\ref{fig:mrnd_latency}~and~\ref{fig:mrnd_iscp} plot our results.
The ping latency (Figure~\ref{fig:mrnd_latency}) increases by around 12ms for
every increase of 1000 branches in the rollback window. The effect
of drift on the completion time of an {\tt iscp} session is less
pronounced (Figure~\ref{fig:mrnd_iscp}).

We also measured failover latencies. In our initial experiments,
a secondary VM takes over as primary within a few seconds of the primary
failure. We intend to perform more detailed experiments measuring failover
latencies in future.

\section{Related Work}
\label{sec:relwork}
Previous efforts on utilizing idle compute capacity include
SETI@Home\cite{seti_at_home}, Folding@Home\cite{folding_at_home},
BOINC project\cite{boinc_project}, etc. Our work has a similar philosophy.
The difference is in the level of abstraction. These previous efforts
require that the programs be written to a specific
programming model, and then provide a middleware which needs to be installed
in all participating host machines. The middleware then coordinates and
schedules the client programs. In contrast, our abstraction is more
general (and often more powerful) than the middleware approach. Our computation
units are VMs, allowing a client full freedom to run her favourite
OS and applications on the participating hosts, without compromising security
and reliability. To provide reliability, the middleware-based approaches
usually constrain the
programming model for easy restartability. In contrast, we allow a completely
flexible programming model and provide reliability through efficient recording
and replaying.

Our current prototype can efficiently record and replay a uniprocessor
VM. We have also implemented record/replay for multiprocessor VMs.
Multiprocessor
record/replay is significantly harder due to the presence of race conditions
on shared memory by multiple processors. We have implemented a page-ownership
scheme based on CREW (concurrent read exclusive write) protocol\cite{smprevirt}
to record and replay a guest OS. We can successfully replay an unmodified
guest, albeit at high overheads. The overheads depend on the workload and could
be as high as 2-3x slowdowns for 2-processor VMs. Another approach,
DoublePlay\cite{doubleplay}, has been proposed to make multiprocessor
record/replay more performant. DoublePlay works by recording the order of
all synchronization operations in the program being recorded. Because a guest
OS could have arbitrary synchronization primitives, it is hard to directly
use DoublePlay's ideas for VM Record/Replay. We are currently working on
approaches to make multiprocessor VM record/replay faster.
\section{Conclusion and Future Work}
\label{sec:conclusion}
We present our initial results on our efforts to develop a community
cloud. We have implemented record/replay inside Linux/KVM and measure its
performance on various workloads. We also analyze concurrent executions of
the record and replay processes to understand their interplay. In future, we
plan to design and test scheduling algorithms for the primary and secondary
replicas of the cloud VMs. We are also working on realizing efficient
multiprocessor VM record/replay.
\section{Acknowledgment}
\label{sec:ack}
The authors would like to thank IBM for supporting this
work in part by grant under the CAS Program.
\bibliographystyle{abbrvnat-custom}
\bibliography{cloud_failover_ccem12}
\end{document}